\documentclass[final,5p,times,twocolumn]{elsarticle}

\usepackage{amssymb}
\usepackage{amsmath}
\usepackage{amsthm}
\usepackage{xcolor}
\usepackage{natbib}
\usepackage{multirow}

\usepackage{hyperref}
\hypersetup{colorlinks=true}
\usepackage{graphicx} 

\begin{document}

\begin{frontmatter}



\title{GBT-SAM: A Parameter-Efficient Depth-Aware Model \\ for Generalizable Brain Tumor Segmentation on mp-MRI}


\author{Cecilia Diana-Albelda$^{a,*}$, Roberto Alcover-Couso$^{a, b}$, $\Acute{A}$lvaro Garc$\acute{\imath}$a-Mart$\acute{\imath}$n$^{a}$, Jesus Bescos$^{a}$, Marcos Escudero-Viñolo$^{a}$} 

\affiliation{organization={Universidad Autónoma de Madrid, Electronics and Communications Technology Dept.},
            city={Madrid},
            postcode={28049}, 
            country={Spain}}

\affiliation{organization={Amazon},
            city={Madrid}, 
            addressline={Spain},
            postcode={28046}, 
            country={This work was done prior to joining Amazon.}}

\begin{abstract}
Gliomas are aggressive brain tumors that require accurate imaging-based diagnosis, where automated segmentation plays a central role in assessing tumor morphology and guiding treatment decisions. Manual delineation of gliomas is time-consuming and prone to variability, motivating the use of deep learning to improve consistency and alleviate clinical workload. However, existing methods often fail to fully exploit the information available in multi-parametric MRI (mp-MRI), particularly inter-slice contextual information. In addition, they typically require considerable computational resources and often lack robustness across different tumor types and imaging conditions. We present GBT-SAM, a parameter-efficient deep learning framework that adapts the large-scale Segment Anything Model (SAM) to volumetric mp-MRI data. GBT-SAM reduces training complexity by selecting fewer than 2.6\% of slices per scan while still incorporating all four MRI modalities, thereby preserving essential tumor-related information at minimal computational cost. Furthermore, our model is trained using a two-step fine-tuning strategy that combines a depth-aware module to capture inter-slice correlations with lightweight LoRA-based adaptation layers, resulting in only 6.5 M trainable parameters, the lowest among existing SAM-based approaches. GBT-SAM is trained exclusively on the BraTS Adult Glioma dataset and achieves a Dice score of 93.02. Also, it is evaluated on additional Meningioma, Pediatric Glioma, and Sub-Saharan Glioma datasets in order to demonstrate its robust generalization performance. These results highlight GBT-SAM as a computationally efficient and domain-robust framework for brain tumor segmentation from mp-MRI. \\ Our code and models are available at \color{blue} \href{https://github.com/vpulab/med-sam-brain}{https://github.com/vpulab/med-sam-brain} \color{black}.
\end{abstract}
\color{black}




\end{frontmatter}



\section{Introduction}
\label{introduction}

\begin{figure}[t]
    \centering
    \includegraphics[width=0.47\textwidth, trim=2cm 3cm 2cm 3cm, clip]{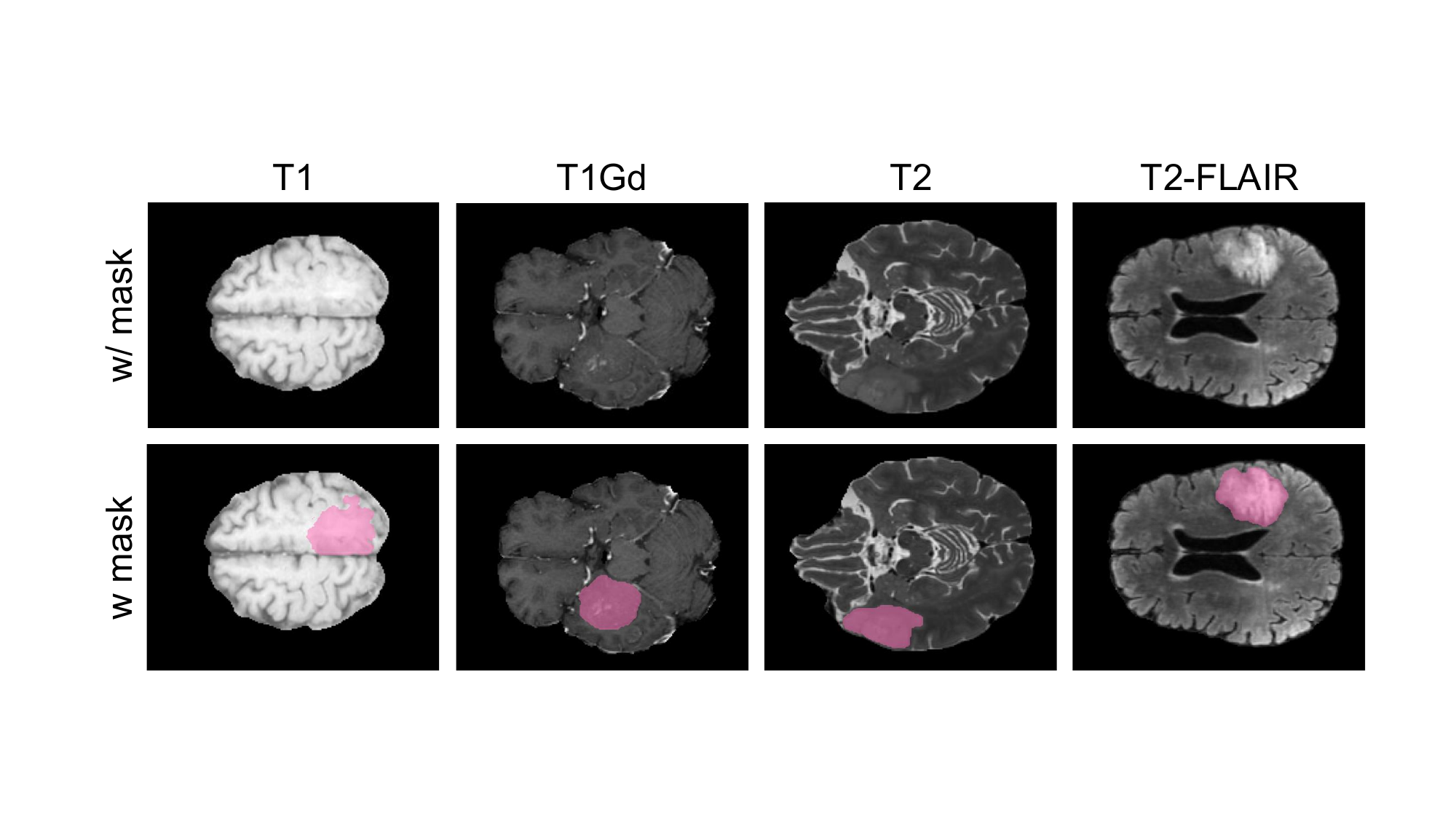}
    \caption{\textbf{MRI modalities.} Examples of tumor visualization in which each column represents a different MRI modality. In these cases, the full tumor extension is not visible in any single modality, highlighting the necessity of leveraging all of them to accurately predict and segment the tumor region. The top row shows the raw MRI images, while the bottom one includes the ground-truth tumor region.}
    \label{fig:why-4-ims}
\end{figure}

Gliomas, a type of primary brain tumor, rank among the most lethal forms of cancer, with significantly lower survival rates than many other malignancies \cite{glioma}. Accurate segmentation of these tumors is critical for diagnosis, enabling precise identification and delineation of the affected region \cite{diana2025review}. Currently, this task is performed manually by radiologists, which is both time-consuming and labour-intensive, ultimately increasing workload and delaying detection. Therefore, an automated mechanism to assist healthcare professionals in segmentation is essential for streamlining the process \cite{brats}.

Deep learning has driven significant advances in automated tumor segmentation over the past decade, offering robust solutions in various medical imaging applications \cite{soni2024multiencoder, zhang2021me}. In particular, Large Vision Models (LVMs) \cite{lvm,alcover2024vlms}, originally trained on extensive natural image datasets, have recently been explored for transfer to medical segmentation tasks. Among these, the Segment Anything Model (SAM) \cite{sam} stands out due to its strong generalization capabilities and its interactive nature, which allows experts to leverage it as an assistance tool accelerating their workflow while maintaining clinical oversight. However, current segmentation frameworks often share three key limitations: 

\textbf{First}, most segmentation models are designed to process three-channel colour images \cite{acc-sam-mi, nnunet, sam-for-mi}. This presents a specific limitation for glioma imaging, where the standard acquisition protocol involves multi-parametric Magnetic Resonance Imaging (mp-MRI) comprising four distinct modalities: T1-weighted (T1), T2-weighted (T2), gadolinium-enhanced T1 (T1c), and T2-FLAIR (Fluid-Attenuated Inversion Recovery) \cite{brats,brats-men,brats-ped,brats-ssa}. Each sequence provides complementary information about tumor structure and pathology, making it crucial to utilize all of them for an accurate delineation \cite{mri-glioma}. To meet model input requirements, it is common to replicate a single modality across three channels. However, this approach severely limits the capabilities of the model, as no single modality captures the full extension of the tumor \cite{cvprw} (see Figure \ref{fig:why-4-ims}).

\textbf{Second}, while MRI captures a 3D brain volume as a sequence of slices, LVMs process each slice independently as an isolated image \cite{sam}. This procedure may be sufficient for some domains, but it ignores inter-slice continuity that radiologists rely on when evaluating tumor extent. To address this shortcoming, certain methods integrate attention mechanisms specifically designed to model these relationships, albeit at the cost of increased computational complexity \cite{medsa, cvprw}.

\textbf{Third}, LVMs generally originate from natural image datasets and therefore lack the specialized knowledge required for medical imaging tasks \cite{alcover2024vlms, cvprw, zero-shot-sam}. As a result, domain adaptation strategies are necessary to bridge the gap between general-purpose training data and the unique characteristics of medical imaging. Moreover, although these models process slices independently, training on all slices from a volume at once can lead them to rely on fixed positional patterns, rather than clinically relevant features, hence reducing generalizability across tumor types and imaging protocols. 

To overcome the above-mentioned limitations, we propose a framework that leverages all four mp-MRI modalities, models inter-slice relationships. Furthermore, we adopt a parameter-efficient fine-tuning strategy based on Low-Rank Adapters (LoRA) to enable effective domain adaptation with minimal computational overhead. Our proposed framework, Generalizable Brain Tumor SAM (GBT-SAM), includes the following contributions:

\begin{enumerate}
    \item \textbf{Multi-modal Adaptation for mp-MRI}: We enable the LVM to process all four mp-MRI modalities (T1, T2, T1c, and T2-FLAIR) jointly, providing the model with the same complementary diagnostic information that clinicians routinely assess. Instead of selecting a single modality and replicating it across channels, we modify the patch embedding layer to accommodate a genuine 4-channel input. We further adopt a two-stage training protocol to better disentangle and exploit the specific contribution of each sequence.
   \item \textbf{Depth-Conditioned Correlation Modelling}: To address the challenge of inter-slice correlation, we introduce a Depth-Condition block that conditions feature representations on adjacent slices. Integrated at multiple stages of the architecture, this component efficiently models inter-slice relationships while incurring negligible additional computational cost.
    \item \textbf{Parameter-Efficient Domain Adaptation}: Recognizing that current models often lack domain-specific knowledge, we implement a parameter-efficient adaptation strategy tailored to clinical imaging tasks. This approach streamlines fine-tuning by learning medically relevant features while minimizing computational overhead.

\end{enumerate}

\begin{figure}[t!]
    \includegraphics[width=0.47\textwidth, trim=2cm 2.5cm 3cm 1.5cm, clip]{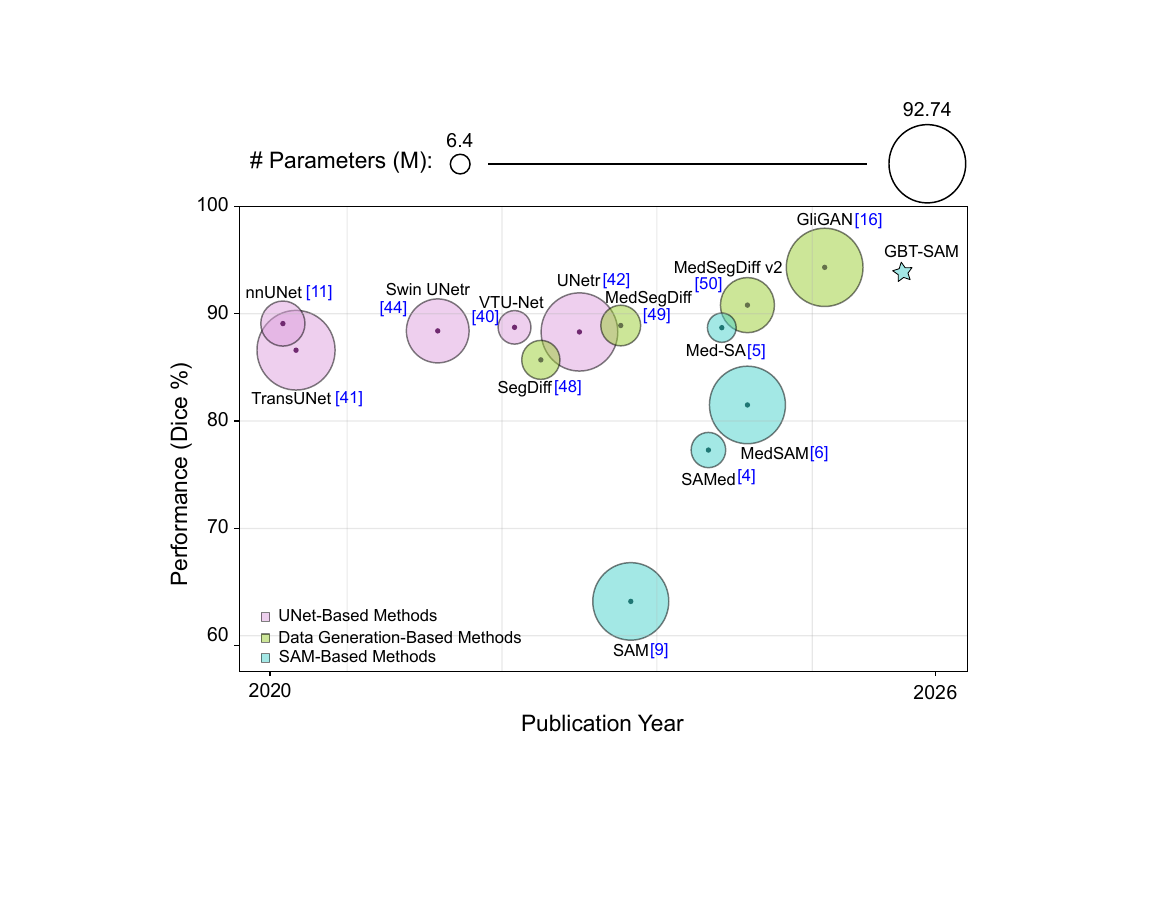}
    \caption{\textbf{Comparison of brain tumor segmentation methods based on performance (Dice Score), publication year, and model size (number of trainable parameters in millions)}. Pink, green and blue bubbles represent UNet-based, Generative, and SAM-based methods respectively. Our method, marked with a star, almost reaches the highest obtained Dice Score while training the smallest number of parameters, highlighting its efficiency and effectiveness compared to state-of-the-art approaches.}
    \label{fig:bubble-plot}
\end{figure}

Building on these contributions, our approach uses up to 14× fewer trainable parameters than the top-performing glioma segmentation method \cite{gligan}, while achieving comparable Dice performance, as illustrated in Figure \ref{fig:bubble-plot}. Additionally, to the best of our knowledge, this is the first SAM-based method evaluated across four distinct brain tumor domains (Adult Glioma, Meningioma, Paediatric Glioma, and Sub-Saharan Gliomas), highlighting its potential generalizability and adaptability to diverse clinical settings.

The content of this paper is organized as follows: Section~\ref{sec:soa} reviews the current state of the art in brain tumor segmentation and LVM adaptation. Section~\ref{sec:method} details our proposed architecture, GBT-SAM, including the multi-modal embedding strategy Depth-Condition block, and the training protocol. Section~\ref{sec:results} presents the experimental setup, ablation studies and both quantitative and qualitative results. Finally, Section~\ref{sec:conclusion} summarizes the main findings, discusses limitations, and outlines directions for future work.

\section{Related Work}
\label{sec:soa}

The most widely used methods for automatic medical imaging segmentation, including brain tumors, focus on three main approaches: UNet-based methods and their variants, data generation techniques i.e. diffusion models and generative adversarial networks (GANs), and foundational models such as SAM-based, which stand out for their interactive segmentation capabilities and greater generalization.

\subsection{UNet-Based Methods}

UNet-based models play a key role in the advancement of medical image segmentation, with ongoing innovations to address evolving challenges. The original UNet \cite{unet} architecture features a contracting path (encoder) to capture context and a symmetric expanding path (decoder) for accurate localization, enhanced by data augmentation to improve generalization. This design is extended in 3D UNet \cite{3dunet}, replacing all 2D operations with 3D counterparts to better segment entire medical image volumes. Recognizing the unique characteristics of medical datasets, nnUNet \cite{nnunet} introduces a self-configurable approach that automatically adjusts the architecture, preprocessing, training, and post-processing for each task. 

Attention mechanisms have further enhanced UNet-based designs. Attention UNet \cite{AttentionUnet} incorporates Attention Gate (AG) modules that automatically focus on relevant structures, improving multi-class segmentation performance. VTUNet \cite{vtunet} extends this concept to 3D medical images, embedding Transformer blocks into the encoder and decoder to capture both local and global features. TransUNet \cite{transunet} introduces transformers as context encoders for 2D images, paired with a UNet-like decoder for spatial information retrieval. This approach is further advanced in UNETR \cite{unetr}, which employs a transformer-based encoder linked to a decoder via skip-connections, enhancing global and multi-scale feature representation for 3D segmentation tasks.
In parallel, the Swin Transformer \cite{Swintransformer} introduces a hierarchical shifted-window self-attention mechanism, which improves computational efficiency by focusing attention on local regions while enabling connections across them. Swin UNETR \cite{swinunetr, kesari2025large} builds on this concept for brain tumor semantic segmentation, utilizing hierarchical transformers to extract features across five resolutions, combined with a Fully-Connected Neural Network (FCNN) decoder through skip-connections. TransDoubleUNet \cite{transdoubleunet} takes this further with a dual U-shaped architecture, integrating dual-scale Swin transformers and a hybrid Convolutional Neural Network (CNN)-transformer decoder to achieve highly accurate segmentation of multimodal brain tumor images. 

Despite their successes, UNet-based enhanced with transformer architectures often struggle to generalize beyond the specific domains for which they are trained. Some studies attempt to overcome this by creating diverse datasets to enhance model adaptability and robustness.

\subsection{Generative Methods}

Data generation-based segmentation methods tackle the challenge of obtaining large, high-quality datasets. Two prominent approaches in this domain are diffusion models and GANs, each offering distinct advancements and contributions.

Diffusion methods, such as De-noising Diffusion Probabilistic Models (DDPMs) \cite{ddpm}, are generative models that use Markov chains to transform simple distributions, such as Gaussian noise, into complex data. The forward diffusion process progressively adds noise to the data, while the inverse process employs an optimized variational framework to reconstruct high-quality data \cite{cvpr-rober}.
SegDiff \cite{segdiff} extends diffusion models for image segmentation by integrating information from the original image with the evolving segmentation map. MedSegDiff \cite{medsegdiff} further adapts DDPMs for medical segmentation by introducing two key components: Dynamic Conditional Encoding (DCE), which adjusts conditions adaptively at each sampling step, and the Feature Frequency Parser (FF-Parser), which removes high-frequency noise. These enhancements improve regional focus and ensure the retention of critical features during the reverse diffusion process.
Advancements like MedSegDiff-v2 \cite{medsegdiffv2} incorporate transformers into the diffusion model framework, resulting in refined segmentation quality.

GANs have also demonstrated effectiveness in generating synthetic data for medical segmentation \cite{liu2025dgeddgan, datta2024brain}. GliGAN \cite{gligan} stands out for employing an architecture with Swin UNETR as the generator and a simple CNN as the discriminator. This model creates synthetic tumors randomly placed in healthy brain regions and combines the synthetic data with real images, improving prediction robustness. Other studies explore novel augmentation methods using multi-armed bandits for dynamic image deformations \cite{xiao2024data}.

Overall, image generation-based models have made notable progress in recent years. However, they still face significant limitations, as the generated images often fall short of the realism compared to real images \cite{alcover2024vlms,alcover2024per,alcover2024gradient}. Additionally, their training and implementation demand substantial computational resources due to their complexity.

\subsection{SAM-Based Methods}

The Segment Anything Model (SAM) \cite{sam} is a foundational model designed to deliver zero-shot segmentation performance across diverse visual domains. Its framework includes a prompt module that provides spatial cues, enabling the segmentation of specific objects within an image. This human-guided segmentation capability is particularly appealing for radiologists, as it allows them to provide initial information about tumor localization that can be further refined, offering a promising pathway for SAM to be integrated as a clinical tool to expedite tumor segmentation. However, as a generalist model, its performance in medical imaging is suboptimal due to the significant differences between natural and medical image domains. Challenges such as low contrast, poorly defined boundaries, and small lesion regions limit its effectiveness in this context \cite{acc-sam-mi, sam-for-mi}. To overcome these limitations, several strategies have been developed to adapt SAM for medical tasks.

To address the domain gap, methods such as SAM-Med2D \cite{sammed2d} and MedSAM \cite{medsam} perform full model fine-tuning using large-scale medical datasets. However, these techniques are computationally expensive and time-consuming, limiting their scalability. In order to reduce computational costs, SAM-U \cite{samu} proposes a no-training method that contrasts predictions obtained from multiple prompts to discard inaccurate ones; however, because the model is not adapted to the target scenario, its performance remains limited. This has motivated the development of more efficient training protocols for SAM adaptation. For instance, SAMed \cite{SAMed} introduces a strategy based on Low-Rank Adaptation (LoRA), fine-tuning the image encoder, prompt encoder, and mask decoder to achieve effective adaptations with fewer resources. In a similar direction, Med-SA \cite{medsa} proposes the Medical SAM Adapter, which incorporates domain-specific medical knowledge through lightweight adaptations.

A further limitation of the original SAM, designed for natural 2D RGB images, is that it does not explicitly account for the volumetric nature of brain tumor imaging. To address this, Med-SA \cite{medsa} introduces the Space-Depth Transpose (SD-Trans) technique, which extends SAM to 3D medical imaging by capturing correlations between volumetric slices. However, this approach comes with significant computational costs, as it relies on processing the entire volumetric image. Alternatively, Medical SAM 2 \cite{medicalsam2} explores the use of SAM 2 \cite{sam2}, a model designed for videos, as an alternative backbone for volumetric data. Medical SAM 2 not only applies segmentation to 3D medical images but also introduces an innovative single-prompt segmentation capability: after an initial prompt, the model can automatically identify and segment similar objects in subsequent slices, optimizing the process. While volumetric images can indeed be treated as analogous to video data, SAM 2 is not fully optimized for such sequences and incurs higher computational costs.

Motivated by the aforementioned limitations, this paper focuses on two key advancements for adapting SAM to mp-MRI brain tumor segmentation: (1) the efficient adaptation of SAM using parameter-efficient fine-tuning (PEFT) tailored to brain tumor imagery, and (2) the introduction of a novel Depth-Condition module that enhances segmentation by integrating information from adjacent upper and lower slices into the decision-making process. In contrast to Med-SA \cite{medsa}, whose SD-Trans module reorders and processes the full 3D volume at once, our approach keeps the original 2D SAM encoder frozen and augments it with lightweight depth-aware conditioning and LoRA-based adapters. This design models inter-slice correlations within small groups of neighbouring slices, avoiding the need to run full 3D self-attention over the entire stack. Similarly, instead of relying on the more complex video-based SAM 2 backbone as in Medical SAM 2 \cite{medicalsam2}, GBT-SAM remains compatible with the original SAM image encoder while achieving volumetric consistency purely through parameter-efficient adapters and the proposed Depth-Condition module. During training, we sample only four slices per volume (approximately $2.6\,\%$ of the full stack) at each iteration to learn these correlations efficiently, whereas at inference time the complete mp-MRI volume is processed to produce a dense 3D prediction.

\section{Method}
\label{sec:method}

 \begin{figure*}[t!]
    \centering
    \includegraphics[width=0.7\textwidth, trim=3cm 5cm 6cm 6cm, clip]{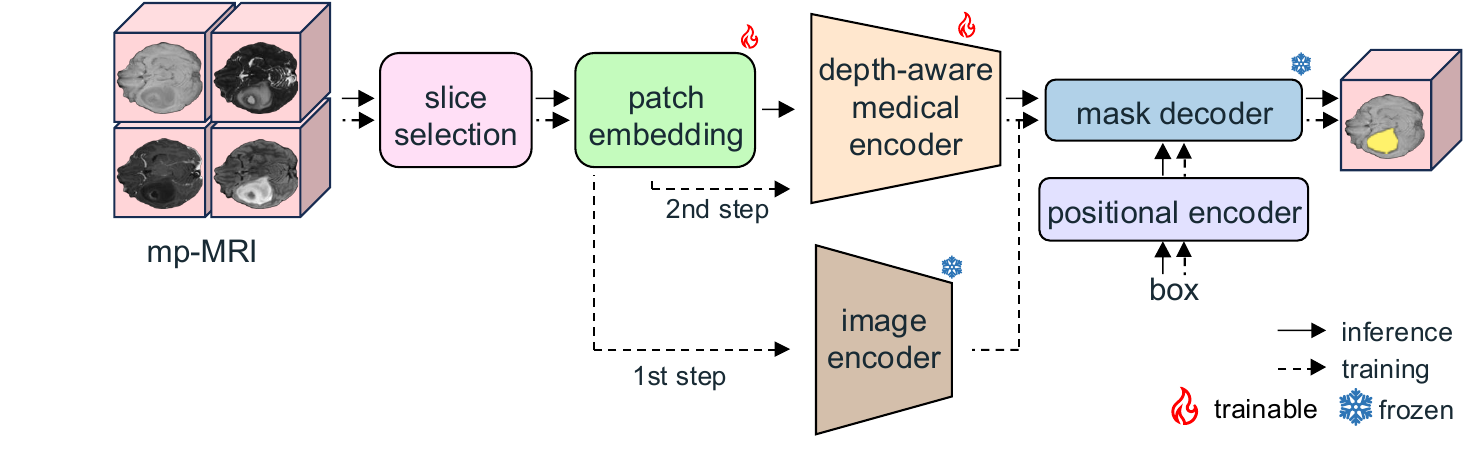}
    \caption{\textbf{GBT-SAM pipeline}. In a first training step, we perform slice selection to reduce computational costs while enhancing generalization capability. Moreover, the patch embedding layer is trained, while the rest of the modules remain frozen: image encoder, responsible for extracting features from the input slices; positional encoder, combining features with the bounding box information; and mask decoder, producing the predicted segmentation. In a second training step, the patch embedding layer is further trained alongside additional trainable components introduced in a modified version of the image encoder (depth-aware medical encoder): LoRA blocks and a Depth-Condition module.}
    \label{fig:2step-training}
    
\end{figure*}

In this section, we first formulate the task of brain tumor segmentation as a volumetric image segmentation framework. Second, we describe the general SAM pipeline which is our building block. Third, we introduce the proposed modifications to the patch embedding to combine information of all mp-MRI modalities, followed by a tailored training setup to effectively adapt the modified layers of SAM. Fourth, we introduce the employed parameter-efficient fine-tuning based on LoRA blocks. Finally, we describe the proposed Depth-Condition block to capture inter-slice correlations. 

Overall, we build upon prior work on adapting SAM to multi-channel MRI, and extend it with a two-stage training scheme and an explicit depth-aware correlation module that are specifically designed for mp-MRI brain tumor segmentation. All together, these contributions describe GBT-SAM's pipeline, illustrated in Figure \ref{fig:2step-training}.

\subsection{Volumetric Image Segmentation}

The volumetric brain tumor segmentation task in this study aims to assign a binary label to each voxel in mp-MRI, indicating whether it belongs to the tumoral region or not \cite{breast-mis, lung-mis}. Consider a mp-MRI volume $X \in \mathbb{R}^{H \times W \times D \times M}$, where \(H\) and \(W\) represent the height and width of each slice, \(D\) is the number of slices, and \(M\) represents the different MRI modalities T1, T2, T1c, and T2-FLAIR.

The goal is to create a model, parametrized by \(\theta\), that assigns a tumor probability to each voxel. We define:

\begin{equation}
f_{\theta}: \mathbb{R}^{H \times W \times D \times M} \to [0,1]^{H \times W \times D},
\end{equation}

such that

\begin{equation}
Y_{\text{pred}} = f_{\theta}(X),
\end{equation}

where \(Y_{\text{pred}} \in [0,1]^{H \times W \times D}\) is the predicted probability of the whole volumes and \(y_{\text{pred}}(h,w,d)\) denotes the probability of a specific voxel being part of the tumor. On the other hand, $Y \in \{0,1\}^{H \times W \times D}$ is the ground-truth segmentation, where \(y(h,w,d) = 1\) indicates that the voxel belongs to the tumor region. Hence, the training problem consists in adjusting \(\theta\) so that \(Y_{\text{pred}}\) closely matches \(Y\).

To quantify the discrepancy between predictions and ground-truth, we employ the Binary Cross-Entropy (BCE) loss \cite{losses}. Although compound losses (e.g. BCE + Dice) are common in medical segmentation, we keep the original BCE loss used in SAM for a fair and stable fine-tuning. The BCE loss can be defined as:

\begin{equation}
\mathcal{L}_{\text{BCE}}(Y, Y_{\text{pred}}) = -\bigl[\,Y \log(Y_{\text{pred}}) + (1 - Y) \log(1 - Y_{\text{pred}})\bigr].
\end{equation}

In this manner, the volumetric tumor segmentation problem is framed as a minimization of the BCE loss function, aiming to find a model that accurately assigns tumor probabilities to each voxel of the mp-MRI volume.

\subsection{SAM Pipeline}

GBT-SAM pipeline builds on top of SAM to perform volumetric brain tumor segmentation, taking as input mp-MRI volumes \(X \in \mathbb{R}^{H \times W \times D \times M}\), as defined above. Given the size and complexity of the input data, a slice sampling strategy is implemented during \emph{training} to reduce computational costs: at each iteration, a random slice \(X_{D}\) is selected together with three additional slices at a fixed distance \(\delta\), as defined in Equation~\ref{slice_selection}, where \(X_{D} \in \mathbb{R}^{H \times W \times M}\) and \(\delta \in \{1,4,10\}\). This approach not only reduces computational requirements by focusing on a limited number of slices per iteration, but also aims to enhance the generalization capabilities of the model by preventing it from relying on complete volume information at once.

\begin{equation}
\label{slice_selection}
   X = \{X_{D- \delta}, X_{D}, X_{D+ \delta}, X_{D+2 \delta}\},
\end{equation}

These selected slices are passed through the patch embedding block, which partitions them into smaller patches \cite{vit}, transforming each into a fixed-dimensional vector representation. The embedded patches are then fed into the ViT blocks of the image encoder, which extracts hierarchical features that capture both local and global contextual information \cite{survey-vit}. These features are enriched with a Depth-Condition block designed to take into account the correlation between slices, hence providing a robust representation of the volumes.

The extracted features are subsequently combined with a bounding box prompt encoding, which supplies spatial guidance regarding the location of the tumor, according to interactive segmentation guidelines of SAM \cite{sam-is}. The combination of contextual and spatial information is processed by the mask decoder to produce a segmentation mask that delineates the final predicted tumor region. 

This pipeline is designed to maximize efficiency by incorporating slice selection and leveraging the architecture to handle volumetric mp-MRI data, promoting accurate tumor segmentation while minimizing computational overhead.

\subsection{Combining All Modalities}

SAM is primarily designed to process RGB images, with its patch embedding layer configured to accept 3-channel inputs. Given that our data consists of 4D mp-MRI, a direct adaptation is necessary to fully utilize this multi-modal volumes.

Existing approaches typically address this limitation by replicating one of the MRI modalities across all three channels \cite{nnunet}, therefore underutilizing the given data and discarding valuable information \cite{survey-mri-brats}. GBT-SAM modifies the patch embedding input layer to accept 4-channel entrances instead of 3 \cite{cvprw}. Formally, let the patch embedding be represented as a function:

\begin{equation}
P: \mathbb{R}^{H \times W \times M} \to \mathbb{R}^{N \times d},
\end{equation}

where \(N\) is the number of patches generated from the input image, $M$ is the number of channels, and \(d\) is the dimensionality of the feature vector for each patch. In the modified architecture, the patch embedding layer processes all four channels simultaneously, with each channel corresponding to one of the MRI modalities. Thus, the input to \(P\) is \(X_D \in \mathbb{R}^{H \times W \times 4}\), where \(X_D\) is a slice selected from the volume. 

This adaptation ensures that the full multi-modal information from mp-MRI is leveraged, mirroring the annotation process used by medical professionals, where each MRI modality provides complementary insights into the tumor structure.

\begin{figure}[t]
    \centering
    \includegraphics[width=0.4\textwidth, trim=4cm 1cm 4cm 1cm, clip]{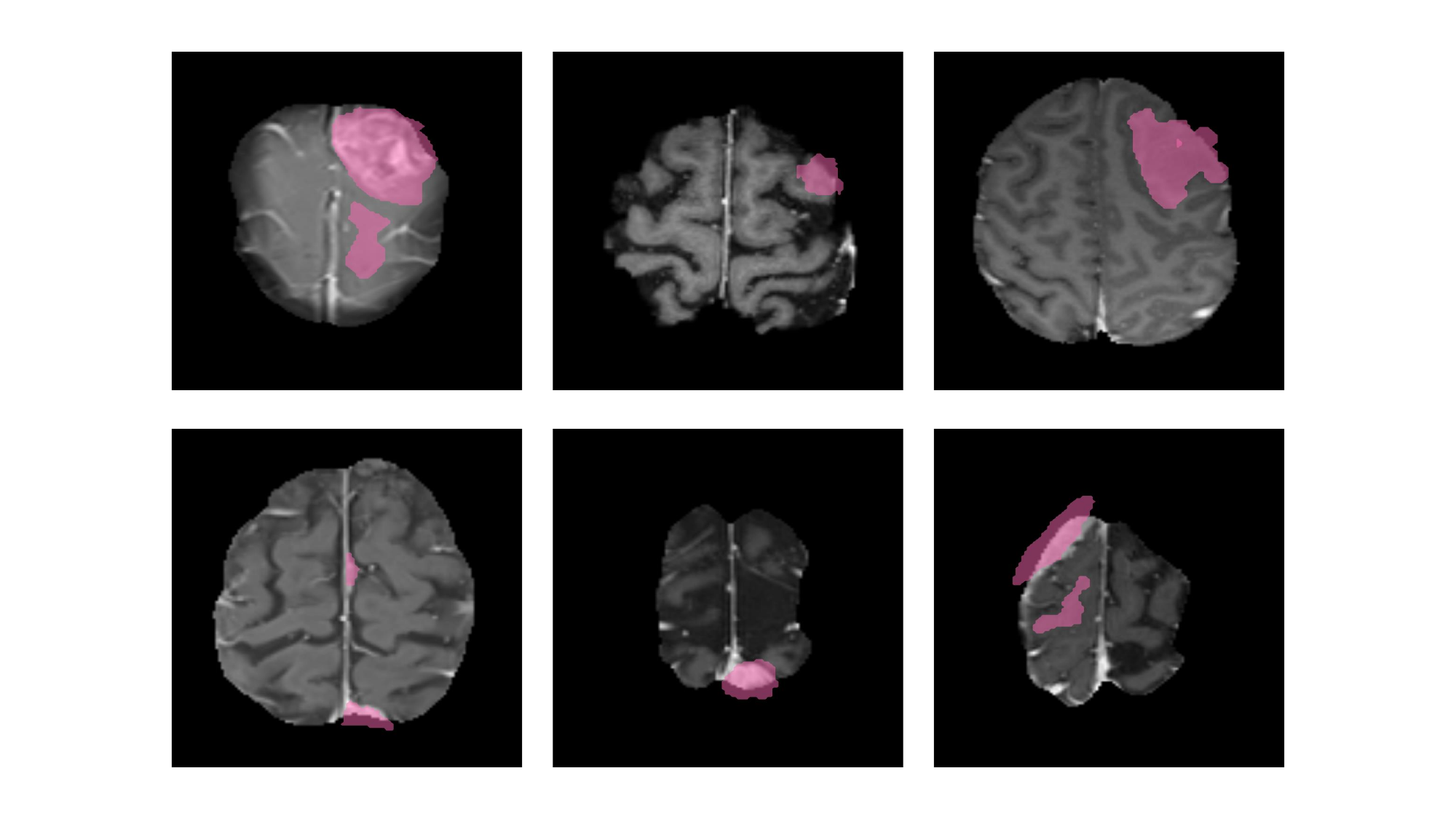}
    \caption{\textbf{Ground-truth incongruity due to slice correlation}. Examples of MRI slices where some pixels are annotated as tumor regions (highlighted in pink) despite the tumor or brain structure is not visible in the pixel. This is due to the fact that doctors may analyse contiguous slices simultaneously, leveraging the volumetric context, hence highlighting the importance of enriching image features by incorporating inter-slice correlations.}
    \label{fig:3d}
\end{figure}

\subsection{Training Setup}

The training of the GBT-SAM follows a two-step protocol to achieve efficient adaptation of SAM to the task of brain tumor segmentation while ensuring optimal utilization of the modified architecture. As the initial layer of the patch embedding is modified and initialized randomly so that it fits $M$ channel inputs, we propose a sequential training to avoid compensatory learning of encoder layers.

In the first step, the entire architecture is frozen except for the patch embedding layer, which processes the full mp-MRI input, capturing information from all four MRI modalities. By training only the patch embedding layer, the model leverages the generalization capabilities of the pre-trained SAM image encoder while adapting to the specific characteristics of the input data, thus ensuring that the patch embedding layer is properly optimized to handle the multi-modal nature of the data before introducing additional trainable parameters. 

In the second step, we employ a parameter-efficient fine-tuning strategy \cite{peft} in which the patch embedding layer remains trainable and LoRA blocks \cite{lora} are introduced within the ViT modules of the image encoder (see Section \ref{subsec: peft}). These LoRA blocks allow the architecture to acquire medical domain-specific knowledge without modifying the entire SAM encoder, thereby significantly reducing the computational cost \cite{lora2, SAMed}. Additionally, a Depth-Condition block is integrated to capture inter-slice correlations (see Section \ref{subsec:depth-condition}).

This two-step training process is designed to ensure a gradual and efficient adaptation of the pipeline. Training the patch embedding layer in isolation during the first step prevents the LoRA blocks from learning to compensate for errors in the patch embedding layer. Consequently, the second step focuses on fine-tuning the system as a whole, achieving a better combination of the image encoder and the modified patch embedding.

\subsection{Parameter-Efficient Fine-Tuning}
\label{subsec: peft}

To efficiently adapt SAM to the medical domain, we employ a Parameter-Efficient Fine-Tuning (PEFT) approach \cite{peft} based on Low-Rank Adaptation (LoRA) \cite{lora}. Consider the set of parameters \(\theta\) that define the segmentation model \(f_{\theta}\). A direct fine-tuning of \(\theta\) over all dimensions would be computationally expensive and could lead to suboptimal performance without large-scale medical data \cite{massive-med-data}. Instead, we decompose the parameter update \(\Delta \theta_l\) for a given layer \(l\) into two low-rank matrices. Being \(\theta_l \in \mathbb{R}^{N \times L}\) the original SAM weights for layer \(l\), we define the updated weights as:

\begin{equation}
\Theta_l = \theta_l + \Delta \theta_l.
\end{equation}

The objective is to constrain \(\Theta_l\) to lie in a low-dimensional subspace. Specifically, we factorize:

\begin{equation}
\Delta \theta_l = A B,
\end{equation}

where \(A \in \mathbb{R}^{N \times r}\) and \(B \in \mathbb{R}^{r \times L}\), with \(r\) controlling the rank of the matrices A and B. 

During training, we keep \(\theta_l\) fixed and only optimize the entries of \(A\) and \(B\). Following \cite{lora}, we initialize \(A \sim \mathcal{N}(0,\sigma)\) and \(B = \mathbf{0}\). The layer \(l\) then produces its output as:

\begin{equation}
\Theta_l(x) = \theta_l x + AB x.
\end{equation}

This factorization restricts the update to a subspace, hence significantly reducing the number of trainable parameters acting as a form of regularization. By doing so, we preserve the foundational structure of SAM while enabling it to specialize to the task of volumetric brain tumor segmentation with minimal additional computational cost. 

\subsection{Depth-Condition Block} 
\label{subsec:depth-condition} 

\begin{figure}[t]
    \centering
    \includegraphics[width=0.5\textwidth, trim=8cm 0.5cm 8cm 0.5cm, clip]{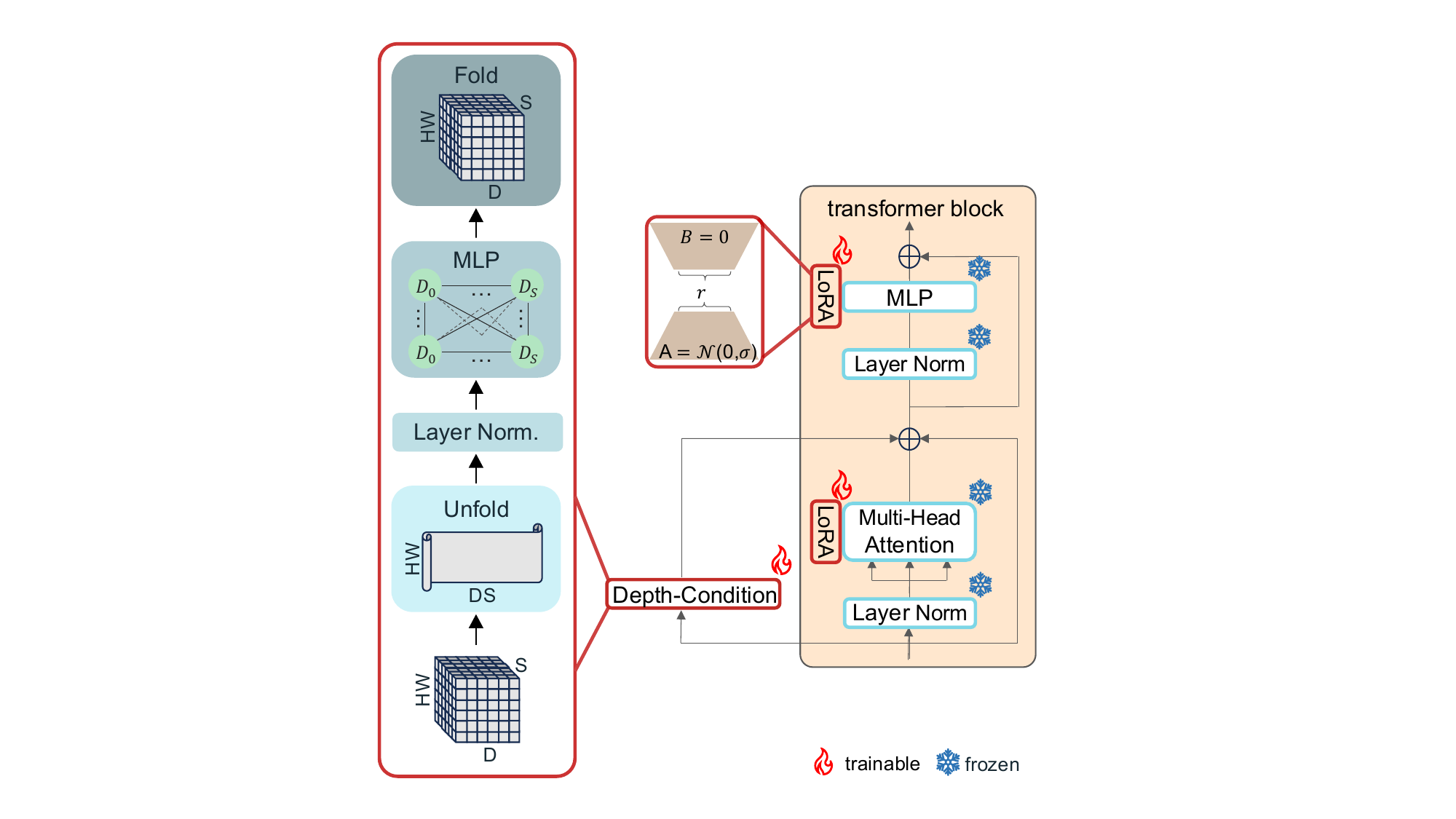}
    \caption{\textbf{Depth-Condition Block.} The Depth-Condition block consists of three main stages: unfolding the volumetric data along the slice dimension with layer normalization, processing it to extract depth-specific features, and folding the data back into the volumetric structure. This Depth-Conditioned output is then integrated into the ViT block, where it complements the features processed by the LoRA modules, enabling the model to leverage both spatial and volumetric information.}
    \label{fig:Depth-Condition}
\end{figure}

When dealing with volumetric medical data, it is crucial to incorporate mechanisms that account for the correlations between features across slices \cite{medsa}. Unlike natural images, where each image is typically processed independently, medical volumes such as mp-MRI provide complementary information distributed throughout the volumetric structure \cite{mri}. As shown in Figure \ref{fig:3d}, there are cases in which tumor regions are annotated in pixels where the tumor is not directly visible, or even where the brain itself is absent, specially for the first or last slices of the volumes. This phenomenon arises because medical experts often annotate tumor regions by simultaneously observing a few slices. Ignoring these correlations by processing slices as independent images can lead to a loss of critical contextual information, thus it is very important that image features capture inter-slice correlations, hence ensuring that the model identifies the dependencies inherent in volumetric data \cite{mri-slice-corr, ct-slice-corr}. 

To address this, we introduce a Depth-Condition module in each ViT block of the image encoder. This module, represented in Figure \ref{fig:Depth-Condition}, operates directly on the volumetric representation \(X \in \mathbb{R}^{H \times W \times D}\) by modelling interactions between slices. 

The block begins by unfolding the volumetric data along the slice dimension \(D\), effectively reshaping the volume into a format that facilitates depth-wise operations while retaining spatial integrity. A layer normalization step is then applied to standardize the input features \cite{gelu}, and then the data is passed through a simple Multi-Layer Perceptron (MLP) designed to learn dependencies and interactions across slices. 

After MLP, the features are folded back into their original volumetric structure, enriched with depth-aware representations that reflect the interactions learned. These Depth-Conditioned features are seamlessly integrated into the ViT blocks in conjunction with LoRA modules, where they complement the spatial information captured by other components of the model. This integration allows the transformer to leverage both spatial and inter-slice dependencies for improved segmentation accuracy.

\section{Experiments \& Results}
\label{sec:results}

In this section, we describe the experimental setup for the study, including dataset characteristics of the different domains, evaluation metrics and computational resources required to perform both training and evaluation phases. Additionally, an extensive evaluation of GBT-SAM has been performed through a detailed ablation study to analyse the impact of significant hyperparameters on the performance of the model. Moreover, we compare our method against current state-of-the-art approaches for brain tumor segmentation, demonstrating its good performance and efficiency. Finally, we provide visual results to illustrate the performance of our model, showcasing predictions alongside ground-truth annotations for a qualitative assessment.

\subsection{Experimental Setup}
\subsubsection{Datasets}

The datasets used in this study have been extracted from the public RSNA-ASNR-MICCAI BraTS 2023 challenge \cite{brats}, which aims to promote research on automatic methods for the diagnosis of gliomas. Our study extends to four datasets corresponding to different brain tumor domains \(D = \{D_1, D_2, D_3, D_4\}\), where:

\begin{itemize}
    \item \(D_1\) (Adult Glioma) \(= \{n_0, n_1, ..., n_{1251}\}\) \cite{brats}
    \item \(D_2\) (Meningioma) \(= \{n_0, n_1, ..., n_{1000}\}\) \cite{brats-men}
    \item \(D_3\) (Pediatric Glioma) \(= \{n_0, n_1, ..., n_{99}\}\) \cite{brats-ped}
    \item \(D_4\) (Sub-Saharan Glioma) \(= \{n_0, n_1, ..., n_{60}\}\) \cite{brats-ssa},
\end{itemize}

being each $n_j$ a patient's full mp-MRI scan. As state-of-the-art methods primarily report their performance on the \(D_1\) dataset, for a fair comparison, we limit our training to \(D_1\). However, we evaluate the generalization capability of GBT-SAM by testing the trained model on all four datasets (\(D_1, D_2, D_3, D_4\)), as well as by including the mean performance on unseen datasets ($DS_{234}$), defined in Equation \ref{mean_dice}. This strategy allows us to assess how effectively the model adapts to diverse brain tumor domains, highlighting its robustness and versatility.

\subsubsection{Evaluation Protocol}

For all experiments, the Dice Score (DS) \cite{dice} is used as the primary evaluation metric, defined as:
\begin{equation}
    DS_i = \frac{2 \cdot |Y_i \cap Y_{\text{pred},i}|}{|Y_i| + |Y_{\text{pred},i}|},
\end{equation}
where \(Y_i\) denotes the ground-truth segmentation for dataset \(D_i\), and \(Y_{\text{pred},i}\) is the corresponding predicted segmentation. The Dice Score is the most widely used metric in brain tumor segmentation and enables consistent and reliable performance comparison across methods \cite{dice2}.

To assess cross-domain generalization, we also compute the mean Dice Score over all unseen datasets as:
\begin{equation}
    \label{mean_dice}
    DS_{234} = \frac{DS_2 + DS_3 + DS_4}{3},
\end{equation}
where \(DS_2\), \(DS_3\), and \(DS_4\) correspond to the Meningioma, Pediatric Glioma, and Sub-Saharan Glioma datasets, respectively.

For all ablation studies reported in this section, we use an internal train–test split (80-20) of the BraTS 2023 Adult Glioma training set. These experiments are designed to analyse the relative impact of each component of GBT-SAM under a controlled setting. In contrast, the results of GBT-SAM presented in Tables~\ref{tab:soa} and~\ref{tab:params-soa} are obtained through the BraTS 2023 online validation set, as evaluated by the official challenge server, ensuring a fair and consistent comparison with existing state-of-the-art methods.

\textbf{Computational Resources.}
Each training session was executed on a single 48GB NVIDIA A40 GPU, utilizing a batch size of 4 and Adam optimizer \cite{adam}.  

\subsection{Ablation Study}

An extensive ablation study has been conducted to analyse the impact of various hyperparameters on the performance of GBT-SAM. By doing this, we aim to maximize the segmentation accuracy and generalization capabilities. 

All experiments in the ablation study are evaluated based on the DS for our four available brain tumor domains: \(D_1\) \cite{brats}, \(D_2\) \cite{brats-men}, \(D_3\) \cite{brats-ped} and \(D_4\) \cite{brats-ssa}. This evaluation allows us to assess the adaptability of the proposed method across diverse clinical scenarios and data distributions.

First of all, we explore different values for the rank (\(r\)) of the matrices introduced by the LoRA blocks \cite{lora}, as this component is specifically designed to provide medical domain knowledge to the network \cite{paranjape2025low}. The results of this experiment are summarized in Table \ref{tab:lora}. 

\begin{table}[!ht]
\centering
\resizebox{\linewidth}{!}{
\begin{tabular}{cccccc}
\hline
 & $\textit{DS}_\textit{1}\uparrow$ & $\textit{DS}_\textit{2}\uparrow$ & $\textit{DS}_\textit{3}\uparrow$ & $\textit{DS}_\textit{4}\uparrow$ & $\textit{DS}_\textit{234}\uparrow$ \\ \hline
$\textit{r=4}$  & $82.53 \pm 0.34$ & $80.62 \pm 0.56$ & $79.25 \pm 1.83$ & $78.18 \pm 2.55$ & $79.35 \pm 1.05$ \\ 
$\textit{r=8}$  & $84.13 \pm 0.32$ & $78.79 \pm 0.79$ & $79.44 \pm 1.55$ & $75.43 \pm 2.16$ & $77.89 \pm 1.21$ \\ 
$\textit{r=12}$ & $85.09 \pm 0.48$ & $80.88 \pm 0.67$ & $\mathbf{83.51 \pm 2.13}$ & $\mathbf{79.00 \pm 1.79}$ & $\mathbf{81.13 \pm 1.53}$ \\ 
$\textit{r=16}$ & $\mathbf{86.10 \pm 0.50}$ & $\mathbf{80.91 \pm 0.56}$ & $80.50 \pm 2.58$ & $75.95 \pm 1.27$ & $79.12 \pm 0.97$ \\ \hline
\end{tabular}}
\caption{\textbf{LoRA rank.} Impact of rank \(r\) in the LoRA blocks on DS across Adult Glioma (\(DS_1\)), Meningioma (\(DS_2\)), Pediatric Glioma (\(DS_3\)), and Sub-Saharan Glioma (\(DS_4\)). Increasing \(r\) improves DS for the training set  \(D_1\), with \(r=16\) achieving the best performance. However, \(r=12\) gets better generalization capabilities (\(DS_{234}\)).}
\label{tab:lora}
\end{table}

From the table, we observe that increasing the rank \(r\) generally leads to improved performance for the training domain (\(D_1\)), with \(r=16\) achieving the highest DS on this dataset. However, when evaluating generalization capability across unseen brain tumor domains (\(DS_{234}\)), \(r=12\) emerges as the optimal choice, providing the best overall balance between segmentation accuracy and generalization. Moreover, the difference in performance on \(D_1\) between \(r=16\) and \(r=12\) is not significant, which also justifies our decision to use \(r=12\) as the default rank in subsequent experiments to ensure robustness across diverse brain tumor domains.

Once the network has been equipped with medical-specific knowledge through LoRA blocks, we evaluate the impact of utilizing the complete mp-MRI volume compared to two alternative approaches: replicating a single modality, as commonly done in RGB-based systems, or combining only the three most representative MRI modalities, i.e. those yielding the highest DS among the four available ones.

\begin{table}[!ht]
\centering
\resizebox{\linewidth}{!}{%
\begin{tabular}{cccccc}
\hline
 MRI Mod. & $\textit{DS}_\textit{1}\uparrow$ & $\textit{DS}_\textit{2}\uparrow$ & $\textit{DS}_\textit{3}\uparrow$ & $\textit{DS}_\textit{4}\uparrow$ & $\textit{DS}_\textit{234}\uparrow$  \\ \hline
$\textit{T1}$ & $49.47 \pm 1.22$ & $68.96 \pm 0.84$ & $57.65 \pm 2.73$ & $42.15 \pm 3.68$ & $56.25 \pm 2.12$ \\ 
$\textit{T1c}$ & $71.27 \pm 0.48$ & $76.96 \pm 0.83$ & $58.36 \pm 1.46$ & $51.09 \pm 2.91$ & $62.14 \pm 1.42$ \\ 
$\textit{T2}$ & $81.86 \pm 0.33$ & $73.84 \pm 0.60$ & $73.14 \pm 1.99$ & $64.18 \pm 1.92$ & $70.37 \pm 1.21$ \\
$\textit{T2-FLAIR}$ & $82.62 \pm 0.49$ & $79.83 \pm 0.53$ & $83.45 \pm 0.76$ & $70.18 \pm 1.50$ & $77.95 \pm 0.82$ \\
$\textit{[T1c, T2, T2-FLAIR]}$ & $\mathbf{86.80 \pm 0.49}$ & $79.99 \pm 0.70$ & $77.52 \pm 1.63$ & $76.03 \pm 1.11$ & $77.85 \pm 0.98$ \\
$\textit{All mp-MRI}$ & $85.09 \pm 0.48$ & $\mathbf{80.88 \pm 0.67}$ & $\mathbf{83.51 \pm 2.13}$ & $\mathbf{79.00 \pm 1.79}$ & $\mathbf{81.13 \pm 1.53}$ \\ \hline
\end{tabular}}
\caption{\textbf{1 vs 3 vs 4 MRI Modalities.} Comparison of replicating individual modalities, combining the three most representative ones, and our proposed approach using all four modalities. Our method leveraging the entire mp-MRI obtains second best DS on Adult Glioma (\(DS_1\)), which is the training set, and the best generalization capability across all not seen brain tumor domains (\(DS_{234}\)): Meningioma, Pediatric Glioma, and Sub-Saharan Glioma.}
\label{tab:mri-mod}
\end{table}

From the results shown in Table \ref{tab:mri-mod}, it is evident that relying on individual modalities significantly underperforms compared to using a combination of modalities or the full mp-MRI. While combining three representative modalities achieves competitive results on \(D_1\) (Adult Glioma), it falls short in generalization to other domains, particularly on \(D_3\) (Pediatric Glioma) and \(D_4\) (Sub-Saharan Glioma). Conversely, leveraging all four MRI modalities demonstrates superior generalization capability \(DS_{234}\), while maintaining strong performance on \(D_1\).

These results highlight that the inclusion of all four mp-MRI modalities allows the network to capture a more comprehensive understanding of tumor morphology and characteristics across varying brain tumor domains, hence confirming the importance of fully utilizing the mp-MRI data.

The next step involves defining the slice selection block to optimize how the data is provided to the network. To achieve this, we evaluate two strategies: selecting 4 distinct random slices per iteration or selecting 1 random slice followed by 3 additional ones at a fixed distance $\delta$ (see Equation \ref{slice_selection}). 

\begin{table}[!ht]
\centering
\resizebox{\linewidth}{!}{%
\begin{tabular}{cccccc}
\hline
 & $\textit{DS}_\textit{1}\uparrow$ & $\textit{DS}_\textit{2}\uparrow$ & $\textit{DS}_\textit{3}\uparrow$ & $\textit{DS}_\textit{4}\uparrow$ & $\textit{DS}_\textit{234}\uparrow$ \\ \hline
\textit{Random} & $83.95 \pm 0.13$ & $79.21 \pm 0.41$  & $70.67 \pm 1.56$  & $69.11 \pm 3.19$ & $73.00 \pm 1.32$ \\ 
\textit{$\delta=10$} & $83.85 \pm 0.29$ & $78.62 \pm 0.73$ & $78.56 \pm 0.59$ & $75.60 \pm 2.48$ & $77.60 \pm 1.05$ \\ 
\textit{$\delta=4$} & $85.09  \pm 0.48$ & $80.88 \pm  0.67$ & $\textbf{83.51} \pm 2.13$ & $79.00 \pm 1.79$ & $81.13 \pm 1.53$ \\ 
\textit{$\delta=1$} & $\textbf{88.11} \pm 0.61$ & $\textbf{85.49} \pm 0.48$  & $83.46 \pm 1.36$ & $\textbf{80.60} \pm 3.10$ & $\textbf{83.18} \pm 1.39$ \\ \hline
\end{tabular}}
\caption{\textbf{Slice selection strategies.} We explore taking 4 random slices or 1 random slice combined with other 3 at a fixed distance ($\delta = \{10, 4, 1\}$). DS across Adult Glioma (\(D_1\)), Meningioma (\(D_2\)), Pediatric Glioma (\(D_3\)), and Sub-Saharan Glioma (\(D_4\)) shows that selecting 4 consecutive slices ($\delta = 1$) yields the best performance.}
\label{tab:slice-choice}
\end{table}

As shown in Table \ref{tab:slice-choice}, the architecture achieves the best performance when using slices with $\delta = 1$, surpassing all other configurations. This highlights the importance of feeding the model sequential images, as consecutive slices provide complementary spatial information, thus enhancing the ability to capture inter-slice correlations. Consequently, the choice of $\delta = 1$ is adopted in subsequent experiments to achieve optimal segmentation performance.

\begin{table}[!ht]
\centering
\resizebox{\linewidth}{!}{%
\begin{tabular}{cccccc}
\hline
 & $\textit{DS}_\textit{1}\uparrow$ & $\textit{DS}_\textit{2}\uparrow$ & $\textit{DS}_\textit{3}\uparrow$ & $\textit{DS}_\textit{4}\uparrow$ & $\textit{DS}_\textit{234}\uparrow$ \\ \hline
$\textit{1p}$ & $88.11 \pm 0.61$ & $85.49 \pm 0.48$ & $83.46 \pm 1.36$ & $80.60 \pm 3.10$ & $83.18 \pm 1.39$ \\ 
$\textit{BB-100-75}$ & $90.90 \pm 0.17$ & $\mathbf{91.84 \pm 0.41}$  & $90.77 \pm 1.17$ & $87.24 \pm 1.93$ & $89.95 \pm 0.92$ \\ 
$\textit{BB-75-75}$ & $\mathbf{91.35 \pm 0.34}$ & $91.31 \pm 0.39$ & $\mathbf{91.18 \pm 0.91}$ & $\mathbf{88.59 \pm 2.70}$ & $\mathbf{90.36 \pm 1.09}$ \\ \hline
$\textit{BB-100-100}$ & $92.17 \pm 0.15$ & $92.68 \pm 0.58$ & $91.54 \pm 1.24$ & $90.62 \pm 2.61$ & $91.61 \pm 1.15$  \\ \hline
\end{tabular}}
\caption{\textbf{Prompts Comparison.} Results for 1 point prompt (\(1p\)) and bounding box prompts (\(BB\)) with varying overlap percentages during training and testing (\(BB\)-100-75 and \(BB\)-75-75). With \(BB\)-100-100 indicating the unrealistic upper DS bound across all domains (Adult Glioma, \(D_1\); Meningioma, \(D_2\); Pediatric Glioma, \(D_3\); and Sub-Saharan Glioma, \(D_4\)), we select \(BB\)-75-75 as the most strong and practical option.}
\label{tab:prompts}
\end{table}

As with all state-of-the-art SAM-based methods, which rely on prompts to harness their interactive segmentation capabilities, in Table \ref{tab:prompts} we explore different types of prompts. Initially, we evaluate the use of a single point prompt (\(1p\)) to guide the segmentation process. Subsequently, we investigate bounding box prompts (\(BB\)) with varying overlap percentages with the tumor region. Specifically, we experiment with \(BB\)-100-100, where bounding boxes have 100\% overlap with the tumor during both training and testing; \(BB\)-100-75, where it is maintained 100\% overlap during training but reduced to 75\% for testing; and \(BB\)-75-75, where 75\% is used in both training and testing. 

From the results, it is clear, as expected, that \(BB\)-100-100 achieves the highest performance in all datasets, indicating the upper performance bound for the method. However, this setting is impractical and unrealistic for clinical use, as it assumes perfect annotations. Therefore, we select the next best-performing option, \(BB\)-75-75, which achieves strong and consistent results while aligning more closely with real-world scenarios.

Lastly, we evaluate different training protocols to determine the most effective approach for the network. The three strategies assessed are: (1) training solely the patch embedding layer, to investigate whether this alone provides sufficient medical domain information; (2) a one-step training protocol where the patch embedding layer and LoRA blocks are trained simultaneously; and (3) a two-step training protocol, where the patch embedding layer is trained first, followed by a second step in which it is retrained alongside LoRA blocks. Each of these protocols is further evaluated with and without the inclusion of our proposed Depth-Condition block, which enriches the model by capturing inter-slice correlations.

\begin{table}[!ht]
\centering
\resizebox{\linewidth}{!}{%
\begin{tabular}{ccccccc}
\hline
Training Process & Slice Corr. & $\textit{DS}_\textit{1}\uparrow$ & $\textit{DS}_\textit{2}\uparrow$ & $\textit{DS}_\textit{3}\uparrow$ & $\textit{DS}_\textit{4}\uparrow$ & $\textit{DS}_\textit{234}\uparrow$ \\ \hline \hline
\textit{Patch Embed.} & & $89.70 \pm 0.34$ & $88.59 \pm 0.46$ & $90.74 \pm 0.98$  & $87.70 \pm 1.82$ & $89.01 \pm 0.90$ \\ \hline
\multirow{2}{*}{\textit{1-step}} & w/o D.C & $91.35 \pm 0.34$ & $91.31 \pm 0.39$ & $91.18 \pm 0.91$ & $88.59 \pm 2.70$ & $90.36 \pm 1.09$ \\ 
  & D.C & $92.76 \pm 0.32$ & $91.33 \pm 0.55$  & $92.10 \pm 1.80$  & $90.20 \pm 1.92$  & $91.21 \pm 1.15$ \\ \hline
\multirow{2}{*}{\textit{2-step}} & w/o D.C & $92.10 \pm 0.69$ & $\mathbf{91.88 \pm 0.61}$ & $90.08 \pm 2.50$ & $89.54 \pm 2.09$  & $90.50 \pm 1.47$  \\ 
      & D.C & $\mathbf{93.54 \pm 0.22}$ & $91.87 \pm 0.61$ & $\mathbf{91.85 \pm 1.78}$ & $\mathbf{91.07 \pm 1.21}$ & $\mathbf{91.60 \pm 0.96}$ \\ \hline
\end{tabular}}
\caption{\textbf{Training Strategy.} D.C is related to the use our proposed Depth-Condition block, while w/o D.C represents not using it. The table evaluates: (1) training only the patch embedding layer, (2) one-step training of patch embedding and LoRA blocks together, and (3) two-step training process, first patch embedding and then patch embedding with LoRA blocks. The two-step process with D.C achieves the best DS (\(DS_1\), \(DS_2\), \(DS_3\), \(D_4\) and \(D_{234}\)) across all datasets (Adult Glioma, Meningioma, Pediatric Glioma and Sub-Saharan Glioma).}
\label{tab:trainings}
\end{table}

The results presented in Table \ref{tab:trainings} demonstrate that the two-step training protocol combined with the Depth-Condition block achieves the best performance across all brain tumor studied scenarios (\(D_1\), \(D_2\), \(D_3\), \(D_4\)). This configuration maximizes segmentation performance by ensuring a robust initial adaptation to mp-MRI data during the first step and enhancing the feature representation during the second step. These findings highlight the importance of separating the training stages to prevent compensatory learning and the critical role of the Depth-Condition block in leveraging inter-slice dependencies for improved volumetric segmentation.

\subsection{Comparison with State-of-the-Art}

We compare GBT-SAM with current state-of-the-art approaches for brain tumor segmentation. All evaluated methods are assessed solely in terms of DS on the Adult Glioma dataset (\(DS_1\)), as this is the most extensively studied domain in the literature \cite{dresunet, unetr}. Moreover, to the best of our knowledge, our method is the first to not only address Adult Glioma segmentation but also explore and compare its generalization capability across Meningioma, Pediatric Glioma, and Sub-Saharan Glioma. Consequently, existing state-of-the-art methods do not provide results or comparisons for the additional datasets included in this study. Hence, \(DS_1\) would be the only shared benchmark for a fair comparison.

\begin{table}[!ht]
\centering
\resizebox{0.7\linewidth}{!}{
\begin{tabular}{lcc}
\hline
\textit{Method} & \textit{Algorithm Family} & $\textit{DS}_\textit{1}\uparrow$  \\ \hline \hline
3D UNet \cite{3dunet} & \multirow{9}{*}{\textit{UNet}} & $84.11$  \\ 
Att. UNet \cite{AttentionUnet} &  & $85.57$  \\
DResU-Net \cite{dresunet} &  & $86.60$   \\ 
TransUNet \cite{transunet} &  & $86.60$ \\ 
UNetr \cite{unetr} &  & $88.30$  \\ 
Swin-UNetr \cite{swinunetr} &  & $88.40$  \\
nnUNet \cite{nnunet} &  & $88.50$   \\ 
VTU-Net \cite{vtunet} &  & $88.73$  \\  
TransDoubleU-Net \cite{transdoubleunet} &  & $92.87$ \\ \hline
SegDiff \cite{segdiff} & \multirow{3}{*}{\textit{Data Generation}} & $85.7$   \\ 
MedSegDiff \cite{medsegdiff} &  & $88.90$  \\ 
MedSegDiff v2 \cite{medsegdiffv2} &  & $90.80$ \\ 
GliGAN \cite{gligan} &  & $\textbf{94.32}$    \\ \hline
SAM \cite{sam} & \multirow{7}{*}{\textit{SAM}} & $63.20$  \\ 
SAMed \cite{SAMed} &  & $77.30$  \\ 
SAM-U \cite{samu} &  & $81.0$ \\ 
MedSAM \cite{medsam} &  & $81.50$  \\ 
SAM-Med2D \cite{sammed2d} &  & $82.90$ \\ 
Med-SA \cite{medsa} &  & $88.70$ \\ 
Medical SAM-2 \cite{medicalsam2} &  & $90.50$ \\ \hline
\textit{GBT-SAM} &  & $\textbf{93.02}$  \\ \hline
\end{tabular}}
\caption{\textbf{Comparison of GBT-SAM with state-of-the-art for brain tumor segmentation in terms of DS on the Adult Glioma dataset $(DS_1)$}. Our method achieves the best performance among SAM-based approaches and ranks second overall, demonstrating its effectiveness and competitiveness against existing methods. \textit{We report WT Dice scores as given in each paper; although all are based on BraTS datasets, minor differences in challenge edition and evaluation protocol may remain.}}
\label{tab:soa}
\end{table}

The results of the comparison between GBT-SAM and current state-of-the-art methods for brain tumor segmentation on the Adult Glioma dataset (\(D_1\)) are presented in Table \ref{tab:soa}. Among the promptable SAM-based approaches, GBT-SAM achieves the highest DS (\(93.02\)), significantly surpassing all other models in the same family. This demonstrates the efficacy of our modifications to SAM for adapting it to the specified domain.

Compared to methods based on UNet architectures, GBT-SAM also outperforms all of the models, surpassing even very popular implementations like Swin-UNetr \cite{swinunetr} (\(88.40\)) and nnUNet \cite{nnunet} (\(88.50\)), being only TransDoubleU-Net \cite{transdoubleunet} (\(92.28\)) close in performance within this category. Additionally, GBT-SAM achieves the second top performance with respect to generative methods, with only GliGAN \cite{gligan} beating it (\(94.32\)).

Overall, GBT-SAM ranks second across all tested methods, which validates the design of our proposed pipeline and its enhancements for handling brain tumor segmentation tasks.

\begin{table}[!ht]
\centering
\resizebox{0.7\linewidth}{!}{
\begin{tabular}{lcc}
\hline
\textit{Method} & $\#$ params (M) $\downarrow$ & $\textit{DS}_\textit{1}\uparrow$  \\ \hline \hline
nnUNet \cite{nnunet}   & $30.76$ &   $88.50$\\ 
VTU-Net \cite{vtunet}    & $17.00$ & $88.73$\\  
SegDiff \cite{segdiff}   & $23.00$ & $85.70$ \\ 
MedSegDiff \cite{medsegdiff}  & $25.00$ &   $88.90$ \\ 
MedSegDiff v2 \cite{medsegdiffv2}    & $46.00$ & $90.80$ \\ 
SAMed \cite{SAMed}  & $18.81$ &   $77.30$ \\ 
Med-SA \cite{medsa}     & $13.00$ & $88.70$ \\ 
\textit{GBT-SAM}  & $\textbf{\textit{6.40}}$ & $\textbf{\textit{93.02}}$ \\ \hline
\end{tabular}}
\caption{\textbf{Efficiency vs Performance.} Comparison of GBT-SAM with state-of-the-art for the DS in Adult Glioma dataset (\(DS_1\)) and the number of trainable parameters. This table includes only methods with $<50M$ trainable parameters. Our method achieves the highest DS while requiring the smallest number of trainable parameters, demonstrating superior efficiency and performance. \textit{We report WT Dice scores as given in each paper; although all are based on BraTS datasets, minor differences in challenge edition and evaluation protocol may remain.}}
\label{tab:params-soa}
\end{table}

To further analyse the efficiency of segmentation methods, we also provide a comparison of GBT-SAM with state-of-the-art approaches that utilize fewer than 50 million trainable parameters. The results, summarized in Table \ref{tab:params-soa}, provide insights into the trade-off between performance and resources. Moreover, this information is also represented visually in Figure \ref{fig:bubble-plot}.

GBT-SAM significantly outperforms all other methods in terms of DS on the Adult Glioma dataset (\(DS_1\)), achieving  \(93.02\). Furthermore, it accomplishes this with only 6.40 million trainable parameters, the lowest among all evaluated methods. For comparison, MedSegDiff v2 \cite{medsegdiffv2} achieves the next highest DS (\(90.80\)) but requires 46 million parameters, over seven times more than GBT-SAM.

This comparison highlights the superior efficiency of GBT-SAM, achieving the highest segmentation accuracy with the fewest trainable parameters.

\subsection{Examples of Qualitative Results}

\begin{figure}[h!]
    \includegraphics[width=0.47\textwidth, trim=3cm 2cm 3cm 1cm, clip]{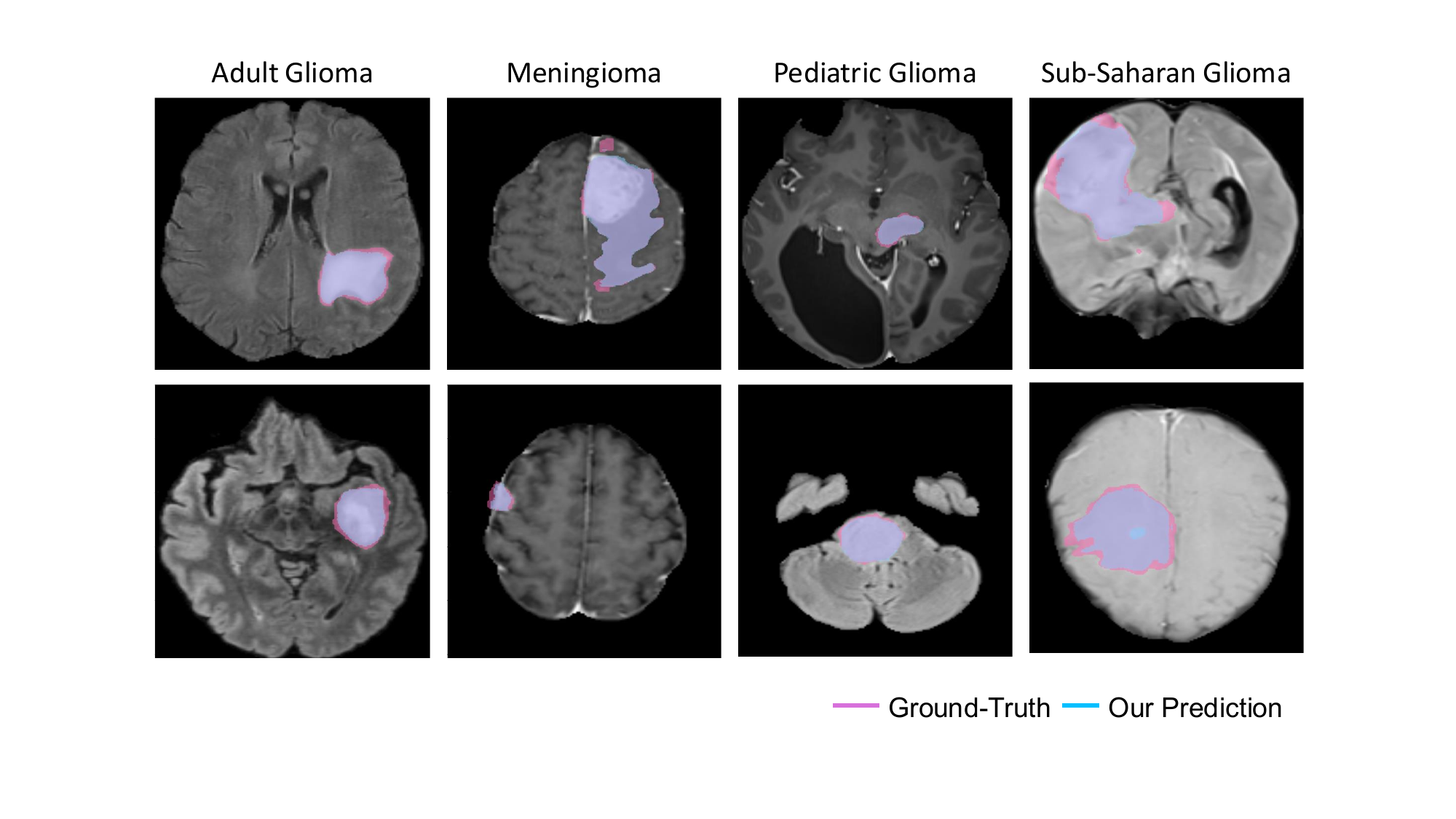}
    \caption{\textbf{Visual comparison of ground-truth (pink) and our predictions (blue)}. The purple regions represent the overlap between the ground-truth and our predictions. Each column shows two examples per domain, demonstrating that the predicted segmentations closely match the annotations across diverse cases.}
    \label{fig:preds}
\end{figure}

The visual results in Figure \ref{fig:preds} show that GBT-SAM produces segmentations that align closely with the ground-truth annotations across all four studied brain tumor domains. The overlap (purple regions) highlights the ability of the model to accurately delineate tumor boundaries and extension. However, there are minor inaccuracies in certain pixels where the model slightly over- or under-segments the tumor regions, particularly in more complex cases or regions with low contrast. These results reflect the robustness of the method in generalizing across different tumor types while acknowledging potential refinement in future work.

\section{Conclusion}
\label{sec:conclusion}

In this work, we introduced GBT-SAM, a framework for brain tumor segmentation that adapts the Segment Anything Model (SAM) to multi-parametric MRI (mp-MRI) data. The proposed architecture combines a restructured patch embedding layer to handle four MRI modalities (T1, T2, T1c, and T2-FLAIR), parameter-efficient fine-tuning with LoRA blocks, and a Depth-Condition block that captures inter-slice correlations in volumetric data. Together, these components enable GBT-SAM to achieve state-of-the-art performance on the BraTS Adult Glioma dataset, with a Dice score of \(93.02\), while maintaining strong generalization to additional tumor domains, including Meningioma, Paediatric Glioma, and Sub-Saharan Glioma.

A key advantage of GBT-SAM is its favourable trade-off between accuracy and computational cost. By jointly exploiting all four MRI modalities, using LoRA-based adaptation, and training with a slice selection strategy that processes fewer than 2.6\% of slices per iteration (4 out of 155), the model reaches competitive performance with fewer than 6.5M trainable parameters. The Depth-Condition block further enhances encoder features by modelling inter-slice dependencies, which is particularly important for volumetric medical imaging, without introducing substantial computational overhead. These characteristics make GBT-SAM a practical candidate for integration into resource-constrained clinical imaging workflows.

Despite these strengths, the current design of GBT-SAM presents some limitations. First, the method relies on bounding box prompts, following interactive segmentation guidelines. While this choice aligns with realistic clinical use, it still requires user-provided annotations at inference time. Exploring strategies for unsupervised or automatically generated prompts is therefore an important direction to reduce annotation burden. Second, the modified patch embedding layer is tailored to four specific MRI modalities, which limits direct applicability to datasets with different or fewer imaging channels, such as BTCV \cite{btcv} or Fundus \cite{ocular} imaging datasets. Additional work is needed to generalize the architecture to variable modality configurations.

Overall, GBT-SAM represents a step towards efficient and generalizable brain tumor segmentation from mp-MRI within the broader context of imaging systems and technology. Future research may focus on extending the framework to support flexible modality inputs, mitigating dependence on manual prompts, and assessing its utility in other anatomical regions and clinical imaging tasks.

\paragraph{\textbf{Acknowledgement}} This work was supported by the Ministerio de Ciencia e Innovación of the Spanish Government (grant PID2021-125051OB-I00) and by the Regional Government of Madrid of Spain (grant TEC 2024/COM-322). 

\paragraph{\textbf{Privacy Considerations}}

Since our data originates from an international challenge, all datasets have been anonymized and preprocessed to ensure privacy and eliminate concerns related to image registration \cite{brats}. This preprocessing phase maintain the data integrity while adhering to ethical and legal standards.




\bibliography{references}

\begin{thebibliography}{10}

\bibitem{glioma}
Michael Weller, Wolfgang Wick, Ken Aldape, Michael Brada, Mitchell Berger, Stefan~M Pfister, Ryo Nishikawa, Mark Rosenthal, Patrick~Y Wen, Roger Stupp, et~al.
\newblock Glioma.
\newblock {\em Nature reviews Disease primers}, 1(1):1--18, 2015.

\bibitem{diana2025review}
Cecilia Diana-Albelda, {\'A}lvaro Garc{\'\i}a-Mart{\'\i}n, and Jesus Bescos.
\newblock A review on deep learning methods for glioma segmentation, limitations, and future perspectives.
\newblock {\em Journal of Imaging}, 11(8):269, 2025.

\bibitem{brats}
Bjoern~H Menze, Andras Jakab, Stefan Bauer, Jayashree Kalpathy-Cramer, Keyvan Farahani, Justin Kirby, Yuliya Burren, Nicole Porz, Johannes Slotboom, Roland Wiest, et~al.
\newblock The multimodal brain tumor image segmentation benchmark (brats).
\newblock {\em IEEE transactions on medical imaging}, 34(10):1993--2024, 2014.

\bibitem{soni2024multiencoder}
Vaibhav Soni, Nikhil~Kumar Singh, Rishi~Kumar Singh, and Deepak~Singh Tomar.
\newblock Multiencoder-based federated intelligent deep learning model for brain tumor segmentation.
\newblock {\em International Journal of Imaging Systems and Technology}, 34(1):e22981, 2024.

\bibitem{zhang2021me}
Wenbo Zhang, Guang Yang, He~Huang, Weiji Yang, Xiaomei Xu, Yongkai Liu, and Xiaobo Lai.
\newblock Me-net: multi-encoder net framework for brain tumor segmentation.
\newblock {\em International Journal of Imaging Systems and Technology}, 31(4):1834--1848, 2021.

\bibitem{lvm}
Jiaqi Wang, Zhengliang Liu, Lin Zhao, Zihao Wu, Chong Ma, Sigang Yu, Haixing Dai, Qiushi Yang, Yiheng Liu, Songyao Zhang, et~al.
\newblock Review of large vision models and visual prompt engineering.
\newblock {\em Meta-Radiology}, page 100047, 2023.

\bibitem{alcover2024vlms}
Roberto Alcover-Couso, Marcos Escudero-Vi{\~n}olo, Juan~C SanMiguel, and Jesus Bescos.
\newblock Vlms meet uda: Boosting transferability of open vocabulary segmentation with unsupervised domain adaptation.
\newblock {\em arXiv preprint arXiv:2412.09240}, 2024.

\bibitem{sam}
Alexander Kirillov, Eric Mintun, Nikhila Ravi, Hanzi Mao, Chloe Rolland, Laura Gustafson, Tete Xiao, Spencer Whitehead, Alexander~C Berg, Wan-Yen Lo, et~al.
\newblock Segment anything.
\newblock In {\em Proceedings of the IEEE/CVF International Conference on Computer Vision}, pages 4015--4026, 2023.

\bibitem{acc-sam-mi}
Sheng He, Rina Bao, Jingpeng Li, P~Ellen Grant, and Yangming Ou.
\newblock Accuracy of segment-anything model (sam) in medical image segmentation tasks.
\newblock {\em arXiv preprint arXiv:2304.09324}, 2023.

\bibitem{nnunet}
Fabian Isensee, Paul~F Jaeger, Simon~AA Kohl, Jens Petersen, and Klaus~H Maier-Hein.
\newblock nnu-net: a self-configuring method for deep learning-based biomedical image segmentation.
\newblock {\em Nature methods}, 18(2):203--211, 2021.

\bibitem{sam-for-mi}
Yuhao Huang, Xin Yang, Lian Liu, Han Zhou, Ao~Chang, Xinrui Zhou, Rusi Chen, Junxuan Yu, Jiongquan Chen, Chaoyu Chen, et~al.
\newblock Segment anything model for medical images?
\newblock {\em Medical Image Analysis}, 92:103061, 2024.

\bibitem{brats-men}
Dominic LaBella, Maruf Adewole, Michelle Alonso-Basanta, Talissa Altes, Syed~Muhammad Anwar, Ujjwal Baid, Timothy Bergquist, Radhika Bhalerao, Sully Chen, Verena Chung, et~al.
\newblock The asnr-miccai brain tumor segmentation (brats) challenge 2023: Intracranial meningioma.
\newblock {\em arXiv preprint arXiv:2305.07642}, 2023.

\bibitem{brats-ped}
Anahita~Fathi Kazerooni, Nastaran Khalili, Xinyang Liu, Debanjan Haldar, Zhifan Jiang, Syed~Muhammed Anwar, Jake Albrecht, Maruf Adewole, Udunna Anazodo, Hannah Anderson, et~al.
\newblock The brain tumor segmentation (brats) challenge 2023: Focus on pediatrics (cbtn-connect-dipgr-asnr-miccai brats-peds).
\newblock {\em arXiv preprint arXiv:2305.17033}, 2023.

\bibitem{brats-ssa}
Maruf Adewole, Jeffrey~D Rudie, Anu Gbdamosi, Oluyemisi Toyobo, Confidence Raymond, Dong Zhang, Olubukola Omidiji, Rachel Akinola, Mohammad~Abba Suwaid, Adaobi Emegoakor, et~al.
\newblock The brain tumor segmentation (brats) challenge 2023: glioma segmentation in sub-saharan africa patient population (brats-africa).
\newblock {\em ArXiv}, 2023.

\bibitem{mri-glioma}
B~POPE Whitney and Garth Brandal.
\newblock Conventional and advanced magnetic resonance imaging in patients with high-grade glioma.
\newblock {\em The quarterly journal of nuclear medicine and molecular imaging: official publication of the Italian Association of Nuclear Medicine (AIMN)[and] the International Association of Radiopharmacology (IAR),[and] Section of the Society of...}, 62(3):239, 2018.

\bibitem{cvprw}
Cecilia Diana-Albelda, Roberto Alcover-Couso, {\'A}lvaro Garc{\'\i}a-Mart{\'\i}n, and Jesus Bescos.
\newblock How sam perceives different mp-mri brain tumor domains?
\newblock In {\em Proceedings of the IEEE/CVF Conference on Computer Vision and Pattern Recognition}, pages 4959--4970, 2024.

\bibitem{medsa}
Junde Wu, Wei Ji, Yuanpei Liu, Huazhu Fu, Min Xu, Yanwu Xu, and Yueming Jin.
\newblock Medical sam adapter: Adapting segment anything model for medical image segmentation.
\newblock {\em arXiv preprint arXiv:2304.12620}, 2023.

\bibitem{zero-shot-sam}
Tassilo Wald, Saikat Roy, Gregor Koehler, Nico Disch, Maximilian~Rouven Rokuss, Julius Holzschuh, David Zimmerer, and Klaus Maier-Hein.
\newblock Sam. md: Zero-shot medical image segmentation capabilities of the segment anything model.
\newblock In {\em Medical Imaging with Deep Learning, short paper track}, 2023.

\bibitem{gligan}
Andr{\'e} Ferreira, Naida Solak, Jianning Li, Philipp Dammann, Jens Kleesiek, Victor Alves, and Jan Egger.
\newblock How we won brats 2023 adult glioma challenge? just faking it! enhanced synthetic data augmentation and model ensemble for brain tumour segmentation.
\newblock {\em arXiv preprint arXiv:2402.17317}, 2024.

\bibitem{unet}
Olaf Ronneberger, Philipp Fischer, and Thomas Brox.
\newblock U-net: Convolutional networks for biomedical image segmentation.
\newblock In {\em Medical image computing and computer-assisted intervention--MICCAI 2015: 18th international conference, Munich, Germany, October 5-9, 2015, proceedings, part III 18}, pages 234--241. Springer, 2015.

\bibitem{3dunet}
{\"O}zg{\"u}n {\c{C}}i{\c{c}}ek, Ahmed Abdulkadir, Soeren~S Lienkamp, Thomas Brox, and Olaf Ronneberger.
\newblock 3d u-net: learning dense volumetric segmentation from sparse annotation.
\newblock In {\em Medical Image Computing and Computer-Assisted Intervention--MICCAI 2016: 19th International Conference, Athens, Greece, October 17-21, 2016, Proceedings, Part II 19}, pages 424--432. Springer, 2016.

\bibitem{AttentionUnet}
Ozan Oktay, Jo~Schlemper, Loic~Le Folgoc, Matthew Lee, Mattias Heinrich, Kazunari Misawa, Kensaku Mori, Steven McDonagh, Nils~Y Hammerla, Bernhard Kainz, et~al.
\newblock Attention u-net: Learning where to look for the pancreas.
\newblock {\em arXiv preprint arXiv:1804.03999}, 2018.

\bibitem{vtunet}
Himashi Peiris, Munawar Hayat, Zhaolin Chen, Gary Egan, and Mehrtash Harandi.
\newblock A robust volumetric transformer for accurate 3d tumor segmentation.
\newblock In {\em International conference on medical image computing and computer-assisted intervention}, pages 162--172. Springer, 2022.

\bibitem{transunet}
Jieneng Chen, Yongyi Lu, Qihang Yu, Xiangde Luo, Ehsan Adeli, Yan Wang, Le~Lu, Alan~L Yuille, and Yuyin Zhou.
\newblock Transunet: Transformers make strong encoders for medical image segmentation.
\newblock {\em arXiv preprint arXiv:2102.04306}, 2021.

\bibitem{unetr}
Ali Hatamizadeh, Yucheng Tang, Vishwesh Nath, Dong Yang, Andriy Myronenko, Bennett Landman, Holger~R Roth, and Daguang Xu.
\newblock Unetr: Transformers for 3d medical image segmentation.
\newblock In {\em Proceedings of the IEEE/CVF winter conference on applications of computer vision}, pages 574--584, 2022.

\bibitem{Swintransformer}
Ze~Liu, Yutong Lin, Yue Cao, Han Hu, Yixuan Wei, Zheng Zhang, Stephen Lin, and Baining Guo.
\newblock Swin transformer: Hierarchical vision transformer using shifted windows.
\newblock In {\em Proceedings of the IEEE/CVF international conference on computer vision}, pages 10012--10022, 2021.

\bibitem{swinunetr}
Ali Hatamizadeh, Vishwesh Nath, Yucheng Tang, Dong Yang, Holger~R Roth, and Daguang Xu.
\newblock Swin unetr: Swin transformers for semantic segmentation of brain tumors in mri images.
\newblock In {\em International MICCAI brainlesion workshop}, pages 272--284. Springer, 2021.

\bibitem{kesari2025large}
Anshika Kesari, Satyajit Maurya, Mohammad~Tufail Sheikh, Rakesh~Kumar Gupta, and Anup Singh.
\newblock Large blood vessel segmentation in quantitative dce-mri of brain tumors: A swin unetr approach.
\newblock {\em Magnetic Resonance Imaging}, page 110342, 2025.

\bibitem{transdoubleunet}
Marjan Vatanpour and Javad Haddadnia.
\newblock Transdoubleu-net: Dual scale swin transformer with dual level decoder for 3d multimodal brain tumor segmentation.
\newblock {\em IEEE Access}, 11:125511--125518, 2023.

\bibitem{ddpm}
Jonathan Ho, Ajay Jain, and Pieter Abbeel.
\newblock Denoising diffusion probabilistic models.
\newblock {\em Advances in neural information processing systems}, 33:6840--6851, 2020.

\bibitem{cvpr-rober}
Pablo Marcos-Manch{\'o}n, Roberto Alcover-Couso, Juan~C SanMiguel, and Jose~M Mart{\'\i}nez.
\newblock Open-vocabulary attention maps with token optimization for semantic segmentation in diffusion models.
\newblock In {\em Proceedings of the IEEE/CVF Conference on Computer Vision and Pattern Recognition}, pages 9242--9252, 2024.

\bibitem{segdiff}
Tomer Amit, Tal Shaharbany, Eliya Nachmani, and Lior Wolf.
\newblock Segdiff: Image segmentation with diffusion probabilistic models.
\newblock {\em arXiv preprint arXiv:2112.00390}, 2021.

\bibitem{medsegdiff}
Junde Wu, Rao Fu, Huihui Fang, Yu~Zhang, Yehui Yang, Haoyi Xiong, Huiying Liu, and Yanwu Xu.
\newblock Medsegdiff: Medical image segmentation with diffusion probabilistic model.
\newblock In {\em Medical Imaging with Deep Learning}, pages 1623--1639. PMLR, 2024.

\bibitem{medsegdiffv2}
Junde Wu, Wei Ji, Huazhu Fu, Min Xu, Yueming Jin, and Yanwu Xu.
\newblock Medsegdiff-v2: Diffusion-based medical image segmentation with transformer.
\newblock In {\em Proceedings of the AAAI Conference on Artificial Intelligence}, volume~38, pages 6030--6038, 2024.

\bibitem{liu2025dgeddgan}
Qiaohong Liu, Weikun Zhang, Yuting Zhang, Xiaoxiang Han, Yuanjie Lin, Xinyu Li, and Keyan Chen.
\newblock Dgeddgan: A dual-domain generator and edge-enhanced dual discriminator generative adversarial network for mri reconstruction.
\newblock {\em Magnetic Resonance Imaging}, 119:110381, 2025.

\bibitem{datta2024brain}
Priyanka Datta and Rajesh Rohilla.
\newblock Brain tumor image pixel segmentation and detection using an aggregation of gan models with vision transformer.
\newblock {\em International Journal of Imaging Systems and Technology}, 34(1):e22979, 2024.

\bibitem{xiao2024data}
Anqi Xiao, Keyi Han, Xiaojing Shi, Jie Tian, and Zhenhua Hu.
\newblock Data augmentation with multi-armed bandit on image deformations improves fluorescence glioma boundary recognition.
\newblock In {\em International Conference on Medical Image Computing and Computer-Assisted Intervention}, pages 130--140. Springer, 2024.

\bibitem{alcover2024per}
Roberto Alcover-Couso, Juan~C SanMiguel, Marcos Escudero-Vinolo, and Pablo Carballeira.
\newblock Per-class curriculum for unsupervised domain adaptation in semantic segmentation.
\newblock {\em The Visual Computer}, pages 1--19, 2024.

\bibitem{alcover2024gradient}
Roberto Alcover-Couso, Marcos Escudero-Vi{\~n}olo, Juan~C SanMiguel, and Jesus Besc{\'o}s.
\newblock Gradient-based class weighting for unsupervised domain adaptation in dense prediction visual tasks.
\newblock {\em arXiv preprint arXiv:2407.01327}, 2024.

\bibitem{sammed2d}
Junlong Cheng, Jin Ye, Zhongying Deng, Jianpin Chen, Tianbin Li, Haoyu Wang, Yanzhou Su, Ziyan Huang, Jilong Chen, Lei Jiang, et~al.
\newblock Sam-med2d.
\newblock {\em arXiv preprint arXiv:2308.16184}, 2023.

\bibitem{medsam}
Jun Ma, Yuting He, Feifei Li, Lin Han, Chenyu You, and Bo~Wang.
\newblock Segment anything in medical images.
\newblock {\em Nature Communications}, 15(1):654, 2024.

\bibitem{samu}
Guoyao Deng, Ke~Zou, Kai Ren, Meng Wang, Xuedong Yuan, Sancong Ying, and Huazhu Fu.
\newblock Sam-u: Multi-box prompts triggered uncertainty estimation for reliable sam in medical image.
\newblock In {\em International Conference on Medical Image Computing and Computer-Assisted Intervention}, pages 368--377. Springer, 2023.

\bibitem{SAMed}
Kaidong Zhang and Dong Liu.
\newblock Customized segment anything model for medical image segmentation.
\newblock {\em arXiv preprint arXiv:2304.13785}, 2023.

\bibitem{medicalsam2}
Jiayuan Zhu, Yunli Qi, and Junde Wu.
\newblock Medical sam 2: Segment medical images as video via segment anything model 2.
\newblock {\em arXiv preprint arXiv:2408.00874}, 2024.

\bibitem{sam2}
Nikhila Ravi, Valentin Gabeur, Yuan-Ting Hu, Ronghang Hu, Chaitanya Ryali, Tengyu Ma, Haitham Khedr, Roman R{\"a}dle, Chloe Rolland, Laura Gustafson, et~al.
\newblock Sam 2: Segment anything in images and videos.
\newblock {\em arXiv preprint arXiv:2408.00714}, 2024.

\bibitem{breast-mis}
Suresh Samudrala and C~Krishna Mohan.
\newblock Semantic segmentation of breast cancer images using densenet with proposed pspnet.
\newblock {\em Multimedia Tools and Applications}, 83(15):46037--46063, 2024.

\bibitem{lung-mis}
Lareib~Fatima Talib, Javaria Amin, Muhammad Sharif, and Mudassar Raza.
\newblock Transformer-based semantic segmentation and cnn network for detection of histopathological lung cancer.
\newblock {\em Biomedical Signal Processing and Control}, 92:106106, 2024.

\bibitem{losses}
Reza Azad, Moein Heidary, Kadir Yilmaz, Michael H{\"u}ttemann, Sanaz Karimijafarbigloo, Yuli Wu, Anke Schmeink, and Dorit Merhof.
\newblock Loss functions in the era of semantic segmentation: A survey and outlook.
\newblock {\em arXiv preprint arXiv:2312.05391}, 2023.

\bibitem{vit}
Alexey Dosovitskiy.
\newblock An image is worth 16x16 words: Transformers for image recognition at scale.
\newblock {\em arXiv preprint arXiv:2010.11929}, 2020.

\bibitem{survey-vit}
Yang Liu, Yao Zhang, Yixin Wang, Feng Hou, Jin Yuan, Jiang Tian, Yang Zhang, Zhongchao Shi, Jianping Fan, and Zhiqiang He.
\newblock A survey of visual transformers.
\newblock {\em IEEE Transactions on Neural Networks and Learning Systems}, 2023.

\bibitem{sam-is}
Yimu Pan, Sitao Zhang, Alison~D Gernand, Jeffery~A Goldstein, and James~Z Wang.
\newblock Ai-sam: Automatic and interactive segment anything model.
\newblock {\em arXiv preprint arXiv:2312.03119}, 2023.

\bibitem{survey-mri-brats}
Tongxue Zhou, Su~Ruan, and Haigen Hu.
\newblock A literature survey of mr-based brain tumor segmentation with missing modalities.
\newblock {\em Computerized Medical Imaging and Graphics}, 104:102167, 2023.

\bibitem{peft}
Zeyu Han, Chao Gao, Jinyang Liu, Jeff Zhang, and Sai~Qian Zhang.
\newblock Parameter-efficient fine-tuning for large models: A comprehensive survey.
\newblock {\em arXiv preprint arXiv:2403.14608}, 2024.

\bibitem{lora}
Edward~J Hu, Yelong Shen, Phillip Wallis, Zeyuan Allen-Zhu, Yuanzhi Li, Shean Wang, Lu~Wang, and Weizhu Chen.
\newblock Lora: Low-rank adaptation of large language models.
\newblock {\em arXiv preprint arXiv:2106.09685}, 2021.

\bibitem{lora2}
Yuchen Zeng and Kangwook Lee.
\newblock The expressive power of low-rank adaptation.
\newblock {\em arXiv preprint arXiv:2310.17513}, 2023.

\bibitem{massive-med-data}
Yu-Cheng Chou, Zongwei Zhou, and Alan Yuille.
\newblock Embracing massive medical data.
\newblock In {\em International Conference on Medical Image Computing and Computer-Assisted Intervention}, pages 24--35. Springer, 2024.

\bibitem{mri}
Suraj~D Serai.
\newblock Basics of magnetic resonance imaging and quantitative parameters t1, t2, t2*, t1rho and diffusion-weighted imaging.
\newblock {\em Pediatric radiology}, 52(2):217--227, 2022.

\bibitem{mri-slice-corr}
Gaoyu Xiao, B~Nicolas Bloch, Jonathan Chappelow, Elizabeth~M Genega, Neil~M Rofsky, Robert~E Lenkinski, John Tomaszewski, Michael~D Feldman, Mark Rosen, and Anant Madabhushi.
\newblock Determining histology-mri slice correspondences for defining mri-based disease signatures of prostate cancer.
\newblock {\em Computerized Medical Imaging and Graphics}, 35(7-8):568--578, 2011.

\bibitem{ct-slice-corr}
Kihwan Choi, Joon~Seok Lim, and Sungwon Kim.
\newblock Self-supervised inter-and intra-slice correlation learning for low-dose ct image restoration without ground truth.
\newblock {\em Expert Systems with Applications}, 209:118072, 2022.

\bibitem{gelu}
Minhyeok Lee.
\newblock Gelu activation function in deep learning: a comprehensive mathematical analysis and performance.
\newblock {\em arXiv preprint arXiv:2305.12073}, 2023.

\bibitem{dice}
Jeroen Bertels, Tom Eelbode, Maxim Berman, Dirk Vandermeulen, Frederik Maes, Raf Bisschops, and Matthew~B Blaschko.
\newblock Optimizing the dice score and jaccard index for medical image segmentation: Theory and practice.
\newblock In {\em Medical Image Computing and Computer Assisted Intervention--MICCAI 2019: 22nd International Conference, Shenzhen, China, October 13--17, 2019, Proceedings, Part II 22}, pages 92--100. Springer, 2019.

\bibitem{dice2}
Tom Eelbode, Jeroen Bertels, Maxim Berman, Dirk Vandermeulen, Frederik Maes, Raf Bisschops, and Matthew~B Blaschko.
\newblock Optimization for medical image segmentation: theory and practice when evaluating with dice score or jaccard index.
\newblock {\em IEEE transactions on medical imaging}, 39(11):3679--3690, 2020.

\bibitem{adam}
Diederik~P Kingma.
\newblock Adam: A method for stochastic optimization.
\newblock {\em arXiv preprint arXiv:1412.6980}, 2014.

\bibitem{paranjape2025low}
Jay~N Paranjape, Shameema Sikder, S~Swaroop Vedula, and Vishal~M Patel.
\newblock Low-rank adaptation of segment anything model for surgical scene segmentation.
\newblock In {\em International Conference on Pattern Recognition}, pages 187--202. Springer, 2025.

\bibitem{dresunet}
Rehan Raza, Usama~Ijaz Bajwa, Yasar Mehmood, Muhammad~Waqas Anwar, and M~Hassan Jamal.
\newblock dresu-net: 3d deep residual u-net based brain tumor segmentation from multimodal mri.
\newblock {\em Biomedical Signal Processing and Control}, 79:103861, 2023.

\bibitem{btcv}
Bennett Landman, Zhoubing Xu, J~Igelsias, Martin Styner, Thomas Langerak, and Arno Klein.
\newblock Miccai multi-atlas labeling beyond the cranial vault--workshop and challenge.
\newblock In {\em Proc. MICCAI Multi-Atlas Labeling Beyond Cranial Vault—Workshop Challenge}, volume~5, page~12, 2015.

\bibitem{ocular}
N~Balakrishna, MB~Mukesh Krishnan, E~Venkata~Ram Sai, S~Vinish Ranganath, K~Sonika, and L~Gouri Priyanka.
\newblock Ocular disease recognition using efficientnet.
\newblock In {\em 2024 3rd International Conference on Applied Artificial Intelligence and Computing (ICAAIC)}, pages 1069--1074. IEEE, 2024.

\end{thebibliography}



\end{document}